
%
\magnification=\magstep1
%
\font\titlefont=cmr10 at 18pt

\font\namefont=cmti10 at 16pt
\font\headfont=cmbx10 at 12pt
\font\subheadfont=cmbx10 at 10pt
\font\subsubheadfont=cmti10

%
%
\newcount\Headernumber \Headernumber=0
%
%
\def\sectskip{\vskip 10pt plus 0pt minus 6pt}
\def\medsectskip{\vskip 5pt plus 0pt minus 3pt}
\def\smallsectskip{\vskip 3pt plus 0pt minus 1pt}
%
%
\def\section#1{\par\sectskip\nobreak%
   \noindent{\headfont
#1}\smallsectskip\nobreak\noindent\ignorespaces}
\def\subsection#1{\par\medsectskip\nobreak%
   \noindent{\subheadfont #1}%
       \smallsectskip\noindent\ignorespaces}
\def\subsubsection#1{\par\smallsectskip \nobreak%
  \noindent{\subsubheadfont #1}
      \smallsectskip\nobreak\noindent\ignorespaces}
\def\headline{\ifnum\Headernumber=0{}\else\ifodd\pageno\tenit%
  \title\hfil\tenrm\folio
   \else\tenrm\folio\hfil\tenit\author\rm\fi\fi}
%
%
%
\def\ref#1{{\goodbreak\parindent=0pt\parskip=0pt \hangindent=10pt
  \hangafter=1 #1.\smallskip\goodbreak}}
\def\title#1{{\it #1}.}
\def\booktitle#1{{\it #1}.}
\def\jtitle#1{{\rm #1},}
%
%

%
\def\caption#1{\vbox{\vskip12pt$${\hbox{\vbox%
{\hsize=128truemm\noindent#1}}}$$\vskip-12pt}}
%
%
\frenchspacing
\voffset=2\baselineskip
\def\maketitle{
{{\noindent\baselineskip=20pt\titlefont\title\par}\vskip20pt
{{\noindent\baselineskip=16pt\namefont\author\par}
}}}
\nopagenumbers
\hfuzz=1.5pt
\vfuzz=1.5pt
\vsize=225truemm
\hsize=138truemm
\parskip=0pt plus 1pt
\parindent=10pt
\baselineskip=12pt
%
%
\def\footline{{\baselineskip=18pt\ifnum\Headernumber=0
{\line{\centerline{\rm\folio}}}\else{}\fi\noindent}}
\output={
  \shipout\vbox{
    \ifnum\Headernumber=0
{\vbox to18pt{\hbox to\hsize{\hfill}\vfill}}
     \vbox to\vsize{\boxmaxdepth\maxdepth\pagecontents}
          {\vbox to18pt{\vfil{\hbox to\hsize{\footline}}}}
    \else{\nopagenumbers\vbox to18pt{\hbox to\hsize{\headline}\vfil}}
               \vbox to\vsize{\boxmaxdepth\maxdepth\pagecontents}
      {\vbox to18pt{\hbox to\hsize{\hfill}\vfill}}
     \fi
}
\global\advance\pageno1\global\advance\Headernumber by 1
\ifnum\outputpenalty>-1000000 \else \dosupereject\fi}


\def\author{Luca Lusanna}
\def\title{Hamiltonian Constraints and Dirac's Observables}

\medsectskip
\medsectskip
\centerline{\bf{Hamiltonian Constraints and Dirac's Observables:}}
\centerline{\bf{from Relativistic Particles towards Field Theory and
General Relativity}}
\sectskip
\sectskip
\sectskip
\sectskip
\centerline {\bf Luca Lusanna}
\smallsectskip
\centerline {Sezione INFN di Firenze}
\centerline {Largo E.Fermi 2 (Arcetri)}
\centerline {50125 Firenze, Italy}
\centerline {e-mail: LUSANNA@FI.INFN.IT}

\sectskip
\sectskip
\sectskip
\sectskip
\sectskip
\sectskip\sectskip\sectskip\sectskip\sectskip\sectskip\sectskip\sectskip
\sectskip\sectskip\sectskip\sectskip
\sectskip\sectskip\sectskip\sectskip

Talk given at the Workshop "Geometry of Constrained Dynamical Systems"
inside the programme on "Geometry and Gravity", June 15-18, 1994,
Newton Institute, Cambridge, UK.

\vfill\eject

\sectskip

Presymplectic manifolds underlie all relevant physical theories, since all
them are, or may be, described by singular Lagrangians [1] and therefore by
Dirac-Bergmann constraints [2] in their Hamiltonian description. In
Galilean physics both Newtonian mechanics [3] and gravity [4] have been
reformulated in this framework. In particular one obtains a multi-time
formulation of non-relativistic particle systems, which generalizes the
non-relativistic limit of predictive mechanics [5] and helps to understand
features unavoidable at the relativistic level, where each particle, due
to manifest Lorentz covariance, has its own time variable. Instead, all both
special and general relativistic theories are always described by singular
Lagrangians. See the review in Ref.[6] and Ref.[7] for the so-called
multi-temporal method for studying systems with first class constraints
(second class constraints are not considered here).

The basic idea relies on Shanmugadhasan canonical transformations [8],
namely one tries to find a new canonical basis in which all first class
constraints are replaced by a subset of the new momenta [when the Poisson
algebra of the original first class constraints is not Abelian, one speaks
of Abelianization of the constraints]; then the conjugate canonical variables
are Abelianized gauge variables and the remaining canonical pairs are
special Dirac's observables in strong involution with both Abelian constraints
and gauge variables. These Dirac's observables, together with the Abelian gauge
variables form a local Darboux basis for the presymplectic manifold [9]
defined by the first class constraints (maybe it has singularities) and
coisotropically embedded in the ambient phase space when there is no
mathematical pathology. In the multi-temporal method each first class
constraint
is raised to the status of a Hamiltonian with a time-like parameter
describing the associated evolution (the genuine time describes the evolution
generated by the canonical Hamiltonian, after extraction from it of the
secondary and higher order first class constraints): in the Abelianized form
of the constraints these "times" coincide with the Abelian gauge variables on
the solutions of the Hamilton equations. These coupled Hamilton equations are
the multi-temporal equations: their solution describes the dependence of the
original canonical variables on the time and on the parameters of the
infinitesimal gauge transformations, generated by the first class constraints.
Given an initial point on the constraint manifold, the general solution
describes the gauge orbit, spanned by the gauge variables, through that point;
instead the time evolution generated by the canonical Hamiltonian (a first
class quantity) maps one gauge orbit into another one. For each system the
main problems are whether the constraint set is a manifold (a stratified
manifold, a manifold with singularities...), whether the gauge orbits can be
built in the large starting from infinitesimal gauge transformations and
whether the foliation of the constraint manifold (of each stratum of it) is
either regular or singular. Once these problems are understood, one can check
whether the reduced phase space (Hamiltonian orbit space) is well defined
and which is the most convenient set of gauge-fixings able to yield a
realization of it.

Usually Shanmugadhasan canonical transformations are defined only locally,
i.e. they are not defined on a whole domain containing the constraint set
or at least its stratum under investigation. However, since all relevant
physical systems must be defined in Minkowski space-time [with their
non-relativistic limit to be understood as a limit in which the velocity of
light c is very big with respect to all other velocities but finite;
the exact Galilean limit $c\rightarrow \infty$ is a contraction which destroys
the genuine relativistic effects due to the geometry of a space-time with
Lorentzian signature (+,-,-,-)], and since for all them it is assumed that the
Poincar\'e algebra and group are globally implemented, then it turns out
(at least for all systems investigated till now) that there is a special
family of Shanmugadhasan canonical transformations globally defined. To find
it, one has to take into account the Lorentz signature of Minkowski
space-time and the consequent existence of various types of Poincar\'e orbits
with different little groups, and to find canonical variables adapted to the
Poincar\'e group (center-of-mass decompositions), to the geometry of the
Poincar\'e orbits and to the first class constraints. The source of globality
may be traced back to the existence of the momentum map for the Poincar\'e
group action on the constraint manifold (or to its different actions on the
various strata of the constraint manifold as we shall see). By means of these
global Shanmugadhasan canonical transformations one obtains a global symplectic
decoupling of the associated Dirac's observables from the Abelian constraints
and gauge variables: in this way one gets natural realizations of the reduced
phase space without the necessity of adding gauge-fixings [10]. The price
for this naturaliness is that the final canonical Hamiltonian is usually
non-linear and non-local in the new variables and one does not know how to
quantize in the standard ways.

Therefore a research program started with the aim to identify all the
consequences of the following three hypotheses for special relativistic
either pointlike or extended systems: i) Lorentz signature of Minkowski
spacetime; ii) Hamiltonian description with first class constraints
(presymplectic approach); iii) global implementation of the Poincar\'e
group.

With special relativistic systems, for which, by assumption, there is a global
implementation of the Poincar\'e algebra and group, the constraint manifold is
a stratified manifold, disjoint union of strata corresponding to the various
types of Poincar\'e orbits. The main stratum consists of all the
configurations of the system with time-like total four-momentum $P^2 > 0$
(with the further subdivision $P_o > 0$ or $P_o < 0$); then there are secondary
strata corresponding to $P^2=0$ ($P_o > 0$ or $P_o < 0$) and $P_{\mu}=0$
(the stratum $P_{\mu}=0$ is connected with the classical background of the
infrared divergences). The stratum corresponding
to space-like orbits $P^2 < 0$ must be absent not to have tachionic effects.
The geometry of the various orbits is different due to the inequivalent
little groups. To study the strata $P^2=0$ and $P_{\mu}=0$ one has to add to
the
original first class constraints the extra first class ones $P^2\approx 0$ and
$P_{\mu}\approx 0$ respectively. Since on the main stratum $P^2 > 0$, one has
the form $W^2=-P^2\, {\vec {\tilde S}}^2$ of the Pauli-Lubanski Casimir of the
Poincar\'e group, this stratum is divided in the two substrata ${\vec {\tilde
S}}\not= 0$ and ${\vec {\tilde S}}=0$, corresponding to the two inequivalent
kinds of orbits of the rotation group generated by the rest-frame Thomas spin
${\vec {\tilde S}}$ (again one has to add ${\vec {\tilde S}}\approx 0$ to the
original constraints to describe this substratum). In the stratum $P^2=0$, when
the little group is O(2), one has $W^{\mu}=\lambda P^{\mu}$, $lim_{P^2
\rightarrow 0}\, W^2/P^2=-\lambda^2$, where $\lambda =\vec P\cdot \vec J/P_o$
is the Poincar\'e invariant helicity; again one must
distinguish $\lambda =0$ from $\lambda \not= 0$.

Let us now review the one-, two- and many-body systems of relativistic
mechanics
, which have been described with first class constraints (see Refs.[11] for the
bibliography; only the papers of the Firenze group will be quoted, since all
them are formulated with a homogeneous formalism). The importance of
relativistic mechanics stems from the fact that quantum field theory has no
particle interpretation: this is forced on it by means of asymptotic states
which, till now, correspond to the quantization of independent one-body systems
described by relativistic mechanics [or relativistic pseudoclassical mechanics
[12], when one adds Grassmann variables to describe the intrinsic spin].
Besides the scalar particle ($P^2-m^2\approx 0$ or $P^2\approx 0$), one has
control on: i) the pseudoclassical electron [13] ($P_{\mu}\xi^{\mu}-m\xi_5
\approx 0$ or $P_{\mu}\xi^{\mu}\approx 0$, where $\xi^{\mu}, \xi_5$ are
Grassmann variables; $P^2-m^2\approx 0$ or $P^2\approx 0$ are implied; after
quantization the Dirac equation is reproduced);
ii) the pseudoclassical neutrino [14] ($P_{\mu}\xi^{\mu}+{i\over 3}\epsilon
^{\mu\nu\rho\sigma}P_{\mu}\xi_{\nu}\xi_{\rho}\xi_{\sigma}\approx 0$, $P^2
\approx 0$, giving the Weyl particle wave equation $P_{\mu}\gamma^{\mu}(1-
\gamma_5)\psi(x)=0$ after quantization); iii) the
pseudoclassical photon [15] ($P^2\approx 0$, $P_{\mu}\theta^{\mu}\approx 0$,
$P_{\mu}\theta^{{*}\mu}\approx 0$, $\theta^{*}_{\mu}\theta^{\mu}\approx 0$,
where $\theta^{\mu}, \theta^{{*}\mu}$ are a pair of complex Grassmann four-
vectors to describe helicity $\pm 1$; after quantization one obtains the photon
wave equations $\bar \sqcup A^{\mu}(x)=0$, $\partial_{\mu}A^{\mu}(x)=0$;
the Berezin-Marinov Grassmann distribution function allows to recover the
classical polarization matrix of classical light and, in quantization, the
quantum polarization matrix with the Stokes parameters);
iv) the vector particle or pseudoclassical massive photon [16] ($P^2-\mu^2+
(1-\lambda)P_{\mu}\theta^{{*}\mu}P_{\nu}\theta^{\nu}\approx 0$,
$\theta^{*}_{\mu}\theta^{\mu}\approx 0$, which, after quantization, reproduce
the Proca-like wave equation $(\bar \sqcup +\mu^2)A^{\mu}(x)-(1-\lambda )
\partial^{\mu}\partial_{\nu}A^{\nu}(x)=0$).

Among the two-body systems, the most important is the DrozVincent-Todorov-Komar
model [17] with an arbitrary action-at-a-distance interaction instantaneous
in the rest frame as shown by its energy-momentum tensor [18] ($P_i^2-m_i^2+
V(R^2_{\perp})\approx 0$, i=1,2, $R^{\mu}_{\perp}=(\eta^{\mu\nu}-P^{\mu}P^{\nu}
/P^2)R_{\nu}$, $R^{\mu}=x_1^{\mu}-x^{\mu}_2$, $P_{\mu}=P_{1\mu}+P_{2\mu}$).
This
model has been completely understood both at the classical and quantum level
[19] (and references therein). Classically it allowed the discovery of the
following sequence of canonical transformations: i) from
$[x^{\mu}_i,p_{i\mu}]$,
i=1,2, to a generic set of center-of-mass and relative variables $[X^{\mu}=
x_1^{\mu}+x_2^{\mu},P_{\mu},R^{\mu},Q_{\mu}={1\over 2}(p_{1\mu}-p_{2\mu}]$;
ii) since the model is defined only for $P^2 > 0$, one can use the standard
Wigner boost for these orbits $L^{\mu}{}_{\nu}(P,{\buildrel o \over P})=
\epsilon^{\mu}{}_{\nu}(P/\eta \sqrt{P^2})=\eta^{\mu}_{\nu}+2P^{\mu}{\buildrel
o \over P}_{\nu}/P^2-(P^{\mu}+{\buildrel o \over P}^{\mu})(P_{\nu}+{\buildrel
o \over P}_{\nu})/(P+{\buildrel o \over P})\cdot {\buildrel o \over P}$, where
$\eta =sign\, P_o$ and ${\buildrel o \over P}_{\mu}=(\eta \sqrt{P^2};\vec 0)$,
to boost at rest the relative variables $R^{\mu}, Q_{\mu}$: in this way one
gets the canonical basis $[{\tilde X}^{\mu},P_{\mu},\epsilon_R=P\cdot Q/\eta
\sqrt{P^2},T_R=P\cdot R/\eta \sqrt{P^2},{\vec {\tilde R}},{\vec {\tilde Q}}]$,
where ${\vec {\tilde R}},{\vec {\tilde Q}}$ are Wigner spin 1 three-vectors,
but ${\tilde X}^{\mu}$ is not a four-vector; iii) finally, from the center-of-
mass variables $[{\tilde X}^{\mu},P_{\mu}]$ one goes to the canonical basis
$[\epsilon =\eta \sqrt{P^2}, T=P\cdot \tilde X/\eta \sqrt{P^2}=P\cdot X/\eta
\sqrt{P^2}, \vec z=\eta \sqrt{P^2}({\vec {\tilde X}}-\vec P{\tilde X}^o/P_o),
\vec k=\vec P/\eta \sqrt{P^2}]$, where $\vec z$, apart a dimensional factor, is
the canonical non-covariant classical analogue of the Newton-Wigner operator
in presence of spin (orbital angular momentum here). In this final canonical
basis one has: a) the Casimir $P^2$ is in the basis via $\epsilon$ and its
conjugate variable is the center-of-mass time in the rest frame; b) suitable
combinations of the constraints may be written as $\chi_R=\epsilon_R-(m^2_1-
m^2_2)/2\epsilon \approx 0$, $\chi =\epsilon^{-2}[\epsilon^2-M^2_{+}(
{\vec {\tilde R}},{\vec {\tilde Q}})][\epsilon^2-M^2_{-}({\vec {\tilde R}},
{\vec {\tilde Q}})]\approx 0$: the first one determines the relative energy
$\epsilon_R$ (so that the conjugate gauge variable is the relative time $T_R$),
while the other one determines the four branches of the mass spectrum
[$\epsilon
=\pm M_{\rho}=\pm (\sqrt{m^2_1+{\vec {\tilde Q}}^2+V(-{\vec {\tilde R}}^2)}+
\rho \sqrt{m^2_2+{\vec {\tilde Q}}^2+V(-{\vec {\tilde R}}^2)})$, $\rho =\pm$;
for equal masses and $P^2 > 0$ there are only two branches $\epsilon =\pm 2
\sqrt{m^2+{\vec {\tilde Q}}^2+V(-{\vec {\tilde R}}^2)}$, since the two
branches with $\epsilon =0$ have $P^2=0$] and T is the conjugate gauge
variable;
c) therefore our final canonical basis $[\epsilon ,T,\epsilon_R,T_R,\vec z,
\vec k,{\vec {\tilde R}},{\vec {\tilde Q}}]$ is a quasi-Shanmugadhasan one
(by redefining $\epsilon_R$ and T, the constraint $\chi_R\approx 0$ may be put
in the form $\epsilon_R^{'}\approx 0$) with $\vec z, \vec k,{\vec {\tilde R}},
{\vec {\tilde Q}}$ playing the role of Dirac's observables
(with respect to $\chi_R$ and
$T_R$) with the evolution determined by $\chi \approx 0$ [i.e. there are four
different Hamiltonians $\pm M_{\rho}({\vec {\tilde R}},{\vec {\tilde Q}})$ for
the four branches]: in this case, by replacing $\chi$ with $\epsilon -(\pm
M_{\rho}({\vec {\tilde R}},{\vec {\tilde Q}}))\approx 0$ one could find
four Shanmugadhasan bases, one for each branch, and therefore the four sets
of final Dirac's observables (Jacobi data), but in general a constraint of
the kind $\epsilon -H\approx 0$ can be included globally in such bases only if
the dynamics generated by the Hamiltonian H is Liouville integrable.
At the quantum level, while one has a system of coupled integro-differential
Klein-Gordon equations coming from the constraints in the original variables,
one gets coupled differential equations in the final ones (in these
variables one can define a Cauchy problem). The mass spectrum and the
elementary solutions have been found in both the formulations [19] for all
potentials for which a complete set of eigenfunctions is known for the
operator $-{\vec \nabla}^2_R+V(-{\vec {\tilde R}}^2)$. Moreover, in both cases
four scalar products, compatible with both equations (i.e. independent from
T and $T_R$), have been found as generalization of the two existing scalar
products of the Klein-Gordon equation: all of them are non-local even in the
limiting free case and differ among themselves for the sign of the norm of
states on different mass-branches.

The connection with the Bethe-Salpeter equation of the quantized model has been
studied in Ref.[20], where it is shown that the constraint wave function can be
obtained from the Bethe-Salpeter one by multiplication for a delta function
containing the relative energy $\epsilon_R$ to exclude the spurious solutions.

The extension of the model to two pseudoclassical electrons and to an electron
and a scalar has been done in Ref.[21], and the first was used to get good
fits to meson spectra.

While in the literature there are various 3- and n-body non-separable models
with first class constraints, see the talk of Longhi[22] for the difficult
case of 3-body separable first class constraints.

Of particular importance is the canonical transformation $(x^{\mu}_i,p_{i\mu})
\mapsto (x^{\mu},P_{\mu},R^{\mu}_a,Q_{a\mu})$, i=1,..,n, a=1,..,n-1, of
Ref.[23],
which transforms the first class constraints $p_i^2-m_i^2\approx 0$ of n free
scalar particles in $P\cdot Q_a\approx 0$, a=1,..,n-1, and in an overall
mass-shell constraint determining the $2^n$ branches of the mass spectrum
($2^n$ is
a topological number, which is broken when some interaction destroys some mass
gap). In analogy to the two-body case for $P^2 > 0$, there is a
quasi-Shanmugadhasan canonical transformation from $[x^{\mu}_i,p_{i\mu}]$,
i=1,..,n. to the base $[\epsilon =\eta \sqrt{P^2}, T=P\cdot X/\eta \sqrt{P^2},
\epsilon_{Ra}=P\cdot Q_a/\eta \sqrt{P^2}, T_{Ra}=P\cdot R_a/\eta \sqrt{P^2},
\vec z,\vec k,{\vec {\tilde R}}_a,{\vec {\tilde Q}}_a]$, a=1,..,n-1, with
${\vec {\tilde R}}_a,{\vec {\tilde Q}}_a$ Wigner spin 1 vectors: the n-1
constraints $\epsilon_{Ra}\approx 0$ are among the new momenta and their
conjugate variables are the n-1 relative times $T_{Ra}$. It is now under
investigation with Pauri and Lucenti
how to go from the sub-base $[{\vec {\tilde R}}_a,{\vec {\tilde
Q}}_a]$ to a sub-base containing $\tilde S=|\, {\vec {\tilde S}}\, |$, so that
this final canonical basis would contain both the Poincar\'e Casimirs,
besides n-1 constraint variables, and therefore would know completely the
geometry of the orbits with $P^2 > 0$; this n-body kinematics would contain
variables relative not only to the center of mass but also to the total spin
(orbital angular momentum) and could be useful in many fields of physics to
take into account the effects of the Lorentz signature of Minkowski
space-time also at low velocities (quasi-Galilean limit). Another feature
which is going to be clarified is the dependence of the inner Lorentz group
generated by $S^{\mu\nu}$ [$J^{\mu\nu}=L^{\mu\nu}+S^{\mu\nu}$, $L^{\mu\nu}=
X^{\mu}P^{\nu}-X^{\nu}P^{\mu}$] and of the zitterbewegung of
center-of-mass variables from the relative time gauge variables. Next, by
replacing the constraints $p_i^2-m^2_i\approx 0$ with $p_i\cdot \xi_i-m_i
\xi_{5i}\approx 0$ (with $\xi^{\mu}_i,\xi_{5i}$ and $\xi^{\mu}_j,\xi_{5j}$
commuting for $i\not= j$), one should be able to include the intrinsic spin
in the final canonical basis.

Both the open and closed Nambu string, after an initial study with light-cone
coordinates, have been treated [24] along the lines of the two-body model in
the sector $P^2 > 0$. Both Abelian Lorentz scalar constraint and gauge
variables
have been found and globally decoupled, and a redundant set of Dirac's
observables $[\vec z,\vec k,{\vec {\tilde a}}_n]$ has been found. It remains an
open problem whether one can extract a global canonical basis of Dirac's
observables from the Wigner spin 1 vectors ${\vec {\tilde a}}_n$, which satisfy
sigma-model-like constraints; if this basis exists (maybe the previous spin
bases could help in this search), it would define the
Liouville integrability of the Nambu string and would open the way to
quantize it in four dimensions.

When a singular Lagrangian may be reconstructed from the first class
constraints
of relativistic mechanics, it is reparametrization invariant; to this
invariance, via the Noether identities imlied by the second Noether theorem, it
corresponds the mass-shell constraint, which is quadratic in the momentum for
the one-body systems. The meaning of the gauge invariance is that the observer
has the freedom to make any choice of what is the "time" with which evolution
is
described. For many-body systems the other constraints may be chosen linear in
the momenta (like $P\cdot Q_a\approx 0$) and the conjugate variables are the
relative times: again their gauge nature implies the freedom of the observer to
describe the system with any given delay among the pairs of constituents. In
the
continuum case of the Nambu string, half of the gauge variables may be
interpreted as the time and the relative times of the points of the string,
while the other half as the freedom for the observer to define what is the
"longitudinal space". In all cases the gauge freedom is of the kind of general
relativity, in which there is built in the freedom to define what are time and
space. Instead in gauge theories, like electromagnetism and Yang-Mills
theories,
the gauge variables correspond to unobservable degrees of freedom. However in
all cases one interpretates the first class constraints as restrictions on
the Cauchy data of the associated Euler-Lagrange equations: in phase space
Dirac's observables are the independent Cauchy data (or Jacobi data, when the
canonical Hamiltonian vanishes).

After the Nambu string, this methodology has been applied to classical gauge
theories following the pioneering work of Dirac [25] for electromagnetism.
By considering a 3+1 splitting of Minkowski space-time and Yang-Mills theory
for a trivial principal bundle over the fixed-time Euclidean space $R^3$ with a
semisimple compact, connected, simply connected Lie group as structure group,
it was possible to find an exact global Shanmugadhasan canonical transformation
to make a symplectic decoupling [26] of the Abelianization of the Gauss' laws
and of the conjugate Abelian gauge variables from the Dirac's observables,
which turn out to be suitable Lie algebra valued transverse vector gauge
potentials and transverse electric fields like in the electromagnetic case.
This is possible in suitable weighted Sobolev spaces, in which the covariant
divergence is an elliptic operator without zero modes [27] and the Gribov
ambiguity is absent [otherwise it is the source of a further stratification
of the constraint manifold [26] and of the presence of cone over cone
singularities due to the stability subgroups of gauge transformations of
certain gauge potentials and field strengths].
The discovery of the Green function for the covariant
divergence allowed to solve the Gauss law constraints and to find the Green
functions of the Faddeev-Popov operator and the square of the covariant
derivative in the case of transverse gauge potentials. After the construction
of a connection-dependent coordinatization of the trivial principal bundle
based on generalized canonical coordinates of first kind on the fibers, the
multi-temporal equations for the gauge potential were solved: the gauge
potential
is decomposed in a pure gauge background connection (the Maurer-Cartan one-form
on the group of gauge transformations or BRST ghost) and in a gauge-covariant
magnetic gauge potential, whose transversality properties were found by using a
generalized Hodge decomposition for one-forms  based on the BRST
operator interpreted as the vertical derivative on the principal bundle. After
an analoguous decomposition of the electric field into transverse and
longitudinal parts (the latter containing transverse gauge potential,
transverse
electric field and Gauss law contributions), the Dirac's observables are
identified as the restriction to the identity cross-section of the trivial
principal bundle of the transverse gauge potential and transverse electric
field. Also the gauge invariant Dirac's observables (but in this case not
physical observables) of Grassmann-valued fermion fields are determined. The
physical Lagrangian and  Hamiltonian,
and the non-Abelian and topological charges are
obtained in terms of the previous Dirac's observables; the form of the
Lagrangian is obtained by means of an explicit realization of the abstract
Riemannian metric of Mitter and Viallet built by using the found Green
functions. When the structure group is SU(3), one has the classical basis of
quantumchromodynamics (QCD). The fundamental role played by the identity
cross-section in the determination of global Dirac's observables raises the
problem whether such observables exist with non-trivial principal bundles and
whether the theory of quantum anomalies should be rephrased as the theory
of obstructions to the existence of global Dirac's observables; if one
classical theory does not admit these observables, then either it is already
pathological at the classical level or one has to find a physical
interpretation of certain gauge degrees of freedom. One has found also the
classical background of the quantum superselection rules: since one has
Dirac's observables $Q_a$ for the non-Abelian charges, one can define a
classical superselection sector as the subset of Dirac's observables which
have vanishing Poisson brackets with the charges $Q_a$, i.e. as the subset of
scalar observables, and a certain value of the Casimirs like $\sum_aQ^2_a$.
In this way only even functions of the "unobservable" Grassmann-valued
fermionic Dirac's observables are selected. One could think to impose
confinement of elementary fermions in a QCD scheme by adding the extra
first class constraints $Q_a\approx 0$. It is an open problem whether there
is a symplectic structure on the scalar Dirac's observables, so that one could
quantize only a superselection sector instead of applying the superselection
rules after quantization. Having eliminated the gauge degrees of freedom
and taken into account the non-localities implied by the implementation of the
Poincar\'e group, the Haag-Kastler program of "local observables" (i.e.
localized on compact domains) for gauge theories should start from Dirac's
observables. Also the role of the center of the group of gauge transformations
and of the winding number have been discussed. Instead, it is still open the
problem of finding Dirac's observables for the standard model of electroweak
interactions with its Higgs mechanism for symmetry breaking: in it some of the
Gauss laws have to be solved in the momenta of the Higgs fields and not in
the Yang-Mills momenta.

Both the Lagrangian and Hamiltonian are non-local and non-polynomial, but
without singularities in the coupling constant; like in the Coulomb gauge they
are not Lorentz invariant, but the invariance can be enforced on them if one
reformulates the theory for the stratum $P^2 > 0$ on space-like
hypersurfaces (on light-cones for the stratum $P^2=0$) following Dirac [2].
With special relativistic theories, one can restrict the space-like
hypersurfaces to hyperplanes orthogonal to $P_{\mu}$ for $P^2 > 0$
(it could be called a Wigner
foliation of Minkowski space-time): in this way the transverse gauge
potential and electric field become Wigner spin 1 three-vectors and the final
dynamics is governed by the first class constraint $\epsilon -H_P\approx 0$,
with $H_P$ being the physical Lorentz invariant Hamiltonian. Another
byproduct of this construction is the indirect proof of the existence of a
center-of-mass decomposition also for classical field theory: namely there is
a canonical basis containing center-of-mass variables ${\tilde X}^{\mu}$,
$P_{\mu}$ or $\epsilon , T, \vec z, \vec k$ plus an infinite number of relative
variables with Wigner covariance. Even if this basis has still to be
constructed (and even if the independent Dirac's observables from the
transverse ones have not yet been extracted with a control on Euclidean
covariance), the important point is that for all extended special relativistic
systems one arrives at a final canonical basis adapted to the Poincar\'e
orbits and to the first class constraints, which contains Dirac's observables
with Wigner covariance and the non-covariant canonical three-vector $\vec z$.

Now in the literature there are three relevant concepts [28] of
center-of-mass position all coinciding in the rest frame: i) the canonical
non-covariant position (also called center of spin; it is the classical basis
of the Newton-Wigner operator in presence of spin), whose role, apart
dimensions, is played by $\vec z$ in the space of Dirac's observables; ii)
the covariant non-canonical Fokker center of inertia, obtained from i) by
Lorentz transformations; iii) the non-covariant non-canonical Moeller
center of energy. Only the second one defines an intrinsic world line and it
can be shown [28] that all the pseudo-world-lines associated with the other
two positions in all possible reference frames fill a world-tube around the
Fokker center of inertia, whose intrinsic transverse radius is determined by
the Poincar\'e Casimirs of the relativistic system (assumed in an irreducible
Poincar\'e representation with $P^2 > 0$ and $W^2\not= 0$): $\rho =|{\vec
{\tilde S}}| /\sqrt{P^2}c=\sqrt{-W^2}/P^2c$. This classical intrinsic unit of
length, whose quantum counterpart is the Compton wavelength of the
configuration multiplied its total spin, has remarkable properties: i) the
criticism to classical theories based on the quantum argument of pair
production applies only inside the world-tube; ii) the world-tube is the
remnant in flat Minkowski space-time of the energy conditions of general
relativity [as shown by Moeller, if a material body has its radius smaller
than $\rho$, then the classical energy density is not definite positive and the
peripheral rotation velocity is higher than the velocity of light]; iii)
the classical relativistic theory of position measurements faces the following
dilemma which has no analogue in Newtonian theories: a) the measurement of the
canonical position [a Dirac observable independent from the gauge-fixings on
the relative times] is frame-dependent, namely this position cannot be
localized inside the world-tube in a covariant way; b) every other center-of-
mass position variable (maybe canonical and covariant as the original naive
center-of-mass $X^{\mu}$, but very often non-canonical) depends on the relative
times and its measurement acquires meaning only after a choice of
gauge-fixings on them. At the quantum level, the situation is even more
complicated due to the theorems of Hegerfeldt [29], according to which, if
the Newton-Wigner position is a self-adjoint oerator, then nearly all wave
packets will spread in space with a velocity higher than the velocity of light
(only wave packets with special power-like tails, living on the boundary of the
Hilbert space, do not have this pathology). Waiting for better develoed
classical and quantum theories of position measurements, we can conclude that
there are strong indications that the center-of-mass of extended relativistic
systems may not be localized inside the world-tube both at the classical and
quantum levels. It could be suggested that in the quantum theory,
enriching the Heisenberg undetermination relations with this veto, one
obtains an ultraviolet cutoff $c\sqrt{P^2}/S$ for the total energy
in the spirit of Dirac and
Yukawa. Let us remark that the insertion of the spin Poincar\'e Casimir in the
final canonical variables could introduce further non-covariant variables; it
will be interesting to find which is the minimal number and meaning of these
non-covariant canonical variables induced by the adaptation to the geometry
of the Poincar\'e orbits and to the first class constraints.

What is not clear is how this ultraviolet cutoff could be used to quantize
the non-linear and non-local physical electromagnetic and Yang-Mills
Hamiltonians. To try to build a consistent framework with this aim in mind,
let us remark that the standard asymptotic Fock spaces used to give a
particle interpretation to quantum field theory and to build S matrix theory
do not constitute a relativistic Cauchy problem for it: since many-particle
states are tensor products of single free particle states, one asymptotic
free particle can be in the absolute future of the others. One reflex of this
freedom ia given by the spurious solutions of the relativistic two-body
bound-state Bethe-Salpeter equation [12], which are excitations in the relative
energy $\epsilon_R$ conjugate to the relative time $T_R$. Therefore, one should
look for a multi-temporal reformulation of the asymptotic states of quantum
field theory by using a canonical basis like the one discussed previously to
describe modified asymptotic states of n free particles on a space-like
hyperplane orthogonal to their total momentum $P_{\mu}$, when $P^2 > 0$. This
is in the spirit of the Tomonaga-Schwinger [30] formulation of quantum field
theory and generalizations of the non-local scalar products described above
for two scalar particles (and extended to spin 1/2 and spin 1 particles)
should be used in the construction. Formal S matrix theory would be unchanged
with the center-of-mass time T replacing the parameter t (there is no
evolution in the relative times due to the quantization of the n-1
first class constraints $\epsilon_{Ra}\approx 0$), but then one should
formulate
a reduction formalism and a perturbative expansion, in which only the total
energy $\epsilon =\eta \sqrt{P^2}$ propagates: the relative energies $\epsilon
_{Ra}$, a=1,..,n-1, should not propagate to avoid spurious solutions of the
resulting bound-state equations. If this reformulation is possible, one should
get a scheme in which the previous ultraviolet cutoff would be natural and
maybe there could be the possibility of a further extension of asymptotic
states to include permanent bound-states (or only them in theories like QCD).
Naturally, the non-covariance of the canonical, probably not self-adjoint,
center-of-mass oerator corresponding to the Dirac observables $\vec z$ would
show up, especially in an attempt to describe this reformulation in a path
integral approach, whose construction till now heavily relies on the
non-relativistic concept of self-adjoint position operators and on their
eigenstates.

Finally the methodology described above should help to solve the first class
constraints of general relativity either in the old tetrad gravity
formulation [31] or in Ashtekar's approach [32] (where solutions already
exist).
In tetrad gravity, there are 16 configuration variables in each space-time
point and 14 first class constraints in phase space, of which 13 linear in the
momenta (three generators of space diffeomorphisms and other ten satisfying a
Poincar\'e algebra). One should solve these 13 constraints and find a
quasi-Shanmugadhasan canonical basis adapted to them; only for some special
class of globally hyperbolic, globally parallelizable, asymptotically flat
manifolds this should be possible, if techniques similar to those used for the
Yang-Mills Gauss laws will work with space diffeomorphisms. If this can be
accomplished, then there should appear in the canonical basis an "energy
variable" $\epsilon$ (like $\epsilon =\eta \sqrt{P^2}$ for special relativistic
[6~systems) such that the time diffeomorphism constraint, which is quadratic
in
the momenta, could be put in the form $\epsilon -H_P\approx 0$ (like for
covariant Yang-Mills theory), with $H_P$ depending only on the two pairs
of Dirac's observables describing the classical graviton degrees of freedom
in each point of space-time; the gauge variable conjugate to $\epsilon$ should
be the "time variable" for this class of general relativistic manifolds.
The eventual globality of the results and the extension to other monifolds
are completely open problems at this stage. However, now both Yang-Mills theory
with fermions and classical general relativity would be put in the same form
and the natural quantization procedure would be the Schroedinger one. If the
idea  of the ultraviolet cutoff determined by the Poincar\'e Casimirs would
work for special relativistic gauge theories, one could hope to use the
Casimirs of the asymptotic Poincar\'e group to try to regularize quantum
gravity. Finally, one should try to solve the constraints of tetrad gravity
coupled to matter and Yang-Mills fields and try to face the unsolved problem
of how to define elementary particles in the framework of general relativity.

\sectskip
{\bf {REFERENCES}}
\sectskip

\ref{1~ {L.Lusanna, 1979, \jtitle{Nuovo Cimento} {\bf {B52}}, 141.}
      {L.Lusanna, 1990, \jtitle{Phys.Rep.} {\bf {185}}, 1.}
      {L.Lusanna, 1991, \jtitle{Riv. Nuovo Cimento} {\bf{14}}, 1.}
      {L.Lusanna, 1990, \break \jtitle{J.Math.Phys.} {\bf{31}}, 2126.}
      {L.Lusanna, 1990, \jtitle{J.Math.Phys.} {\bf{31}}, 428.}
      {M.\break
      Chaichian, D.Louis Martinez and L.Lusanna, 1993, \jtitle{"Dirac's
      Constrained Systems: The Classification of Second-Class Constraints"},
      Helsinki Univ. preprint HU-TFT-93-5}}\hfill\break
\ref{2~ {P.A.M.Dirac, 1950, \jtitle{Can.J.Math.} {\bf 2},129; \booktitle
      {"Lectures on  Quantum Mechanics"}, Belfer Graduate School of Science,
      Monographs Series, Yeshiva University, 1964.}
      {J.L.Anderson and P.G.Bergmann, 1951, \jtitle{Phys.Rev.} {\bf{83}},
1018.}
      {P.G.Bergmann and J.Goldberg, 1955, \jtitle{Phys.Rev.} {\bf{98}}, 531}}
      \hfill\break
\ref{3~ {G.Longhi, L.Lusanna and J.M.Pons, 1989, \jtitle{J.Math.Phys.}
        {\bf{30}},  1893}}\hfill\break
\ref{4~ {R.De Pietri, L.Lusanna and M.Pauri, 1994, \jtitle{"Standard and
      Generalized Newtonian Gravities as 'Gauge' Theories of the Extended
      Galilei Group: I) The Standard Theory; II) Dynamical Three-Space
      Theories"}, Parma Univ. preprints}}\hfill\break
\ref{5~ {L.Bel, 1970, \jtitle{Ann.Inst.H.Poincar\'e} {\bf{307}}; 1971,
{\bf{14}},
      189; 1973, {\bf{18}},57; 1977, \booktitle{Mecanica Relativista
Predictiva}
      , report UABFT-34, Universidad Autonoma Barcelona (unpublished)}}
      \hfill\break
\ref{6~ {L.Lusanna, \jtitle{"Dirac's Observables: from Particles to Strings
      and Fields"}, talk given at the International Symposium on
      \booktitle{Extended
      Objects and Bound States}, Karuizawa 1992, eds. O.Hara, S.Ishida and
      S.Naka, World Scientific, 1993}}\hfill\break
\ref{7~ {L.Lusanna, \jtitle{"Classical Observables of Gauge Theories from the
      Multitemporal Approach"}, talk given at the Conference 'Mathematical
      Aspects of Classical Field Theory', Seattle 1991, in \jtitle{Contemporary
      Mathematics} {\bf{132}}, 531 (1992)}}\hfill\break
\ref{8~ {S.Shanmugadhasan, 1973,\jtitle{J.Math.Phys.} {\bf{14}}, 677.}
      {L.Lusanna, 1993, \break \jtitle{"The Shanmugadhasan
      Canonical Transformation, Function Groups and the Extended Second Noether
      Theorem"}, \jtitle{Int.J.Mod.Phys.} {\bf{A8}}, 4193}}\hfill\break
\ref{9~ {M.J.Gotay, J.M.Nester and G.Hinds,  1978,\jtitle{J.Math.Phys.}
        {\bf{19}}, 2388; M.J.\break
      Gotay and J.M.Nester, 1979,\jtitle{Ann.Inst.Henri Poincar
      $\grave e$} {\bf{A30}}, 129 and 1980, {\bf{A32}}, 1. M.J.Gotay and
      J.$\grave S$niatycki, 1981,\jtitle{Commun.Math.Phys.} {\bf{82}}, 377.
      M.J.Gotay, 1982,\jtitle{Proc.Am.Math.Soc.} {\bf{84}}, 111; 1986, \jtitle
      {J.Math.Phys.} {\bf{27}}, 2051. G.Marmo, N.Mukunda and J.Samuel, 1983,
      \jtitle{Riv.Nuovo Cimento} {\bf 6}, 1}}\hfill\break
\ref{10 {J.M.Arms, M.J.Gotay and G.Jennings, 1990, \jtitle{Adv.Math.}
         {\bf{79}}, 43}}\hfill\break
\ref{11 {F.M.Lev, 1993, \jtitle{Rivista Nuovo Cim.} {\bf{16,n.2}}, 1.
         \booktitle{Relativistic Action-at-a-Distance}, ed. J.A.Llosa,
         Lecture Notes Phys. n.162, Springer, 1982. \booktitle{Constraint's
         Theory and Relativistic Dynamics}, eds. G.Longhi and L.Lusan-\break
         na, World Scientific, 1987.
         L.Lusanna, 1981, \jtitle{Nuovo Cim.}{\bf{65B}},135; 1984, in Proc.
         VII Seminar on \booktitle{Problems of High Energy Physics and Quantum
         Field Theory}, Protvino USSR, vol.I, p.123; in \booktitle{Gauge
         Field Theories}, XVIII Karpacz School, Harwood, 1986}}\hfill\break
\ref{12 {R.Casalbuoni, 1976, \jtitle{Nuovo Cim.} {\bf{33A}}, 115 and 389.
         F.A.Berezin and M.S.Marinov, 1977, \jtitle{Ann.Phys. (N.Y.)} {\bf
         {104}}, 336. A.Barducci, R.Casalbuo-
         ni and L.Lusanna, 1977, \jtitle
         {Nuovo Cim.Lett.} {\bf{19}}, 581; 1977, \jtitle{Nucl.Phys.}
         {\bf{B124}}, 93; 1981, \jtitle{Nucl.Phys.} {\bf{B180}} [FS2],
         141}}\hfill\break
\ref{13 {A.Barducci, R.Casalbuoni and L.Lusanna, 1976, \jtitle{Nuovo Cim.}
         {\bf{35A}}, 377}}\hfill\break
\ref{14 {A.Barducci, R.Casalbuoni, D.Dominici and L.Lusanna, 1981, \break
         \jtitle{Phys.Lett.} {\bf{100B}}, 126}}\hfill\break
\ref{15 {A.Barducci and L.Lusanna, 1983, \jtitle{Nuovo Cim.} {\bf {77A}},
         39}}\hfill\break
\ref{16 {A.Barducci and L.Lusanna, 1983, \jtitle{J.Phys.} {\bf{A16}}, 1993}}
         \hfill\break
\ref{17 {Ph.Droz Vincent, 1969, \jtitle{Lett.Nuovo Cim.} {\bf{1}}, 839;
         1970, \jtitle{Phys.Scr.} {\bf{2}}, 129; 1975, \jtitle{Rep.Math.Phys.}
         {\bf{8}}, 79. I.T.Todorov, 1976, \jtitle{Report Comm. JINR E2-10125,
         Dubna} (unpublished); 1978, \jtitle{Ann.Inst.H.Poincar\'e} {\bf{28A}},
         207. A.Komar, 1978, \jtitle{Phys.Rev.} {\bf{D18}}, 1881 and
         1887}}\hfill\break
\ref{18 {A.Barducci, R.Casalbuoni and L.Lusanna, 1979, \jtitle{Nuovo Cim.}
         {\bf{54A}}, 340}}\hfill\break
\ref{19 {G.Longhi and L.Lusanna, 1986,\jtitle{Phys.Rev.} {\bf{D34}}, 3707}}
         \hfill\break
\ref{20 {H.Sazdjian, 1981, \jtitle{Ann.Phys.(N.Y.)} {\bf{136}}, 136; 1986,
         \jtitle{Phys.Rev.} {\bf{D33}}, 3401; 1987, \jtitle{J.Math.Phys.}
         {\bf{28}}, 2618 and 1988, {\bf{29}}, 1620; 1989, \jtitle{Ann.Phys.
         (N.Y.)} {\bf{191}}, 52; in Proc.Int.Symp. \booktitle{Extended
         Objects and Bound Systems}, eds. O.Hara, S.Ishida and S.Naka, World
         Scientific, 1992. J.Bijtebier and J.Brockaert, 1992, \jtitle{Nuovo
         Cim.} {\bf {A105}}, 351 and 625; in Proc.Int.Symp. \booktitle{Extended
         Objects and Bound Systems}, eds. O.Hara, S.Ishida and S.Naka, World
         Scientific, 1992}}\hfill\break
\ref{21 {H.W.Crater and P.Van Alstine, 1982, \jtitle{J.Math.Phys.} {\bf{23}},
         1697; 1983, \jtitle{Ann.Phys.(N.Y)} {\bf{148}}, 57; 1984, \jtitle
         {Phys.Rev.Lett.} {\bf{53}}, 1577; 1984, \jtitle{Phys.Rev.} {\bf{D30}},
         2585; 1986, {\bf{D34}}, 1932; 1987, {\bf{D36}}, 3007; 1988,
{\bf{D37}},
         1982; 1990, \jtitle{J.Math.Phys.} {\bf{31}}, 1998; 1992, \jtitle{Phys.
         Rev.} {\bf{D46}}, 766. H.W.Crater, R.L.\break
         Becker, C.Y.Wong and P.Van Alstine, 1992, \jtitle{Phys.Rev.}
{\bf{D46}}
         , 5117; in Proc.Int.Symp. \booktitle{Extended
         Objects and Bound Systems}, eds. O.Hara, S.\break
         Ishida and S.Naka, World
         Scientific, 1992. H.W.Crater and D.Yang, 1991, \jtitle{J.Math.Phys}
         {\bf{32}}, 2374}}\hfill\break
\ref{22 {G.Longhi, talk at this Workshop}}\hfill\break
\ref{23 {L.Lusanna, 1981, \jtitle{Nuovo Cim.} {\bf {64A}}, 65}}\hfill\break
\ref{24 {F.Colomo, G. Longhi and L.Lusanna, 1990,\jtitle{Int.J.Mod.Phys.}
         {\bf{A5}}, 3347; 1990, \jtitle{Mod.Phys.Lett.} {\bf{A5}}, 17.
         F.Colomo and L.Lusanna, 1992,\break \jtitle{Int.J.Mod.Phys.}
         {\bf{A7}}, 1705  and 4107}}\hfill\break
\ref{25 {P.A.M.Dirac, 1955, \jtitle{Can.J.Phys.} {\bf{33}}, 650}}\hfill\break
\ref{26 {L.Lusanna, 1994, \jtitle{Dirac's Observables for Classical
         Yang-Mills Theory with Fermions}, Firenze Univ. preprint DFF 201/1/
         94}}\hfill\break
\ref{27 {V.Moncrief, 1979, \jtitle{J.Math.Phys.} {\bf {20}}, 579}}\hfill\break
\ref{28 {M.Pauri, 1980, in \booktitle{Group Theoretical Methods in Physics},
         ed. K.B.Wolf, Lecture Notes Phys. {\bf {135}}, Sprnger, Berlin; 1971,
         \jtitle{Invariant Localization and Mass-Spin Relations in the
         Hamiltonian Formulation of Classical Relativistic Dynamics}, Parma
         Univ. preprint IFPR-T-019 (unpublished)}}\hfill\break
\ref{29 {G.C.Hegerfeldt, 1989, \jtitle{Nucl.Phys. (Proc.Suppl.)} {\bf {B6}},
         231}}\hfill\break
\ref{30 {S.Tomonoga, 1946, \jtitle{Prog.Theor.Phys.} {\bf{1}}, 27; 1948,
         \jtitle{PhysRev.} {\bf{74}}, 224. Z.Koba, T.Tati and S.Tomonoga,
         1947, \jtitle{Prog.Theor.Phys.} {\bf{2}}, 101 and 198. S.Kanesawa
         and S.Tomonaga, 1948, \jtitle{Prog.Theor.Phys.} {\bf{3}}, 1 and
         101.\break J.Schwinger, 1948, \jtitle{Phys.Rev.} {\bf{73}}, 416 and
         {\bf{74}}, 1439}}\hfill\break
\ref{31 {J.E.Nelson and C.Teitelboim, 1978, \jtitle{Ann.Phys. (N.Y.)}
         {\bf{116}}, 86. M.\break
         Henneaux, 1983, \jtitle{Phys.Rev.} {\bf{D27}},
         986. J.M.Charap and J.E.Nelson, 1983, \jtitle{J.Phys.} {\bf{A16}},
         1661 and 3355; 1986, \jtitle{Class.Quantum Grav.} {\bf{3}}, 1061.
         J.M.\break Charap, in \booktitle{Constraint's Theory and Relativistic
         Dynamics}, eds. G.Longhi and L.Lusanna, World Scientific, 1987.
         J.M.Charap, M.Henneaux and J.E.Nelson, 1988, \jtitle{Class.Quantum
         Grav.} {\bf{5}}, 1405}}\hfill\break
\ref{32 {A.Ashtekar, 1987, \jtitle{Phys.Rev.} {\bf{D36}}, 1587; \booktitle{
         New Perspectives in Canonical Gravity}, Bibliopolis,
         1986; \booktitle{Lectures on Non-Perturbative Canonical Gravity},
         World Scientific, 1991. M.Henneaux, J.E.Nelson and C.Schomblond,
         1989, \jtitle{Phys.Rev.} {\bf{D39}}, 434. R.Capovilla, J.Dell and
         T.Jacobson, 1989, \jtitle{Phys.
         Rev.Lett.} {\bf{63}}, 2325. J.D.
         Romano, 1993, \jtitle{Gen.Rel.Grav.} {\bf{25}}, 759}}

\bye